\documentclass[iop]{emulateapj}
\usepackage{natbib}

\newcommand{\zz}{$\zeta$}

\newcommand{\zone}{$\zeta_1$}
\newcommand{\ztwo}{$\zeta_2$}
\newcommand{\pp}{\mathcal{P}}

\newcommand{\Kepler}{{\sl Kepler}\ }
\newcommand{\Keplert}{\emph{Kepler}}
\newcommand{\Keplers}{{\sl Kepler's} }

\newcommand{\be}{\begin{equation}}
\newcommand{\ee}{\end{equation}}
\newcommand{\bea}{\begin{eqnarray}}
\newcommand{\eea}{\end{eqnarray}}

\shorttitle{ Architecture of \Kepler Multiplanets}
\shortauthors{Fabrycky et al.}

\begin{document}
\title{
Architecture of \Keplers Multi-transiting Systems: II. New investigations with twice as many candidates
}
\author{
Daniel C. Fabrycky\altaffilmark{1,2},  
%
Jack J. Lissauer\altaffilmark{3},   
Darin Ragozzine\altaffilmark{4}, 
Jason F. Rowe\altaffilmark{3,5},    
Jason H. Steffen\altaffilmark{6},   
Eric Agol\altaffilmark{7},  
Thomas Barclay\altaffilmark{3,8},  
Natalie Batalha\altaffilmark{3,9},  
William Borucki\altaffilmark{3}, 
David R. Ciardi\altaffilmark{10}, 
Eric B. Ford\altaffilmark{11,12,13}, 
Thomas N. Gautier\altaffilmark{14}, 
John C. Geary\altaffilmark{4}, 
Matthew J. Holman \altaffilmark{4},  
Jon M. Jenkins \altaffilmark{3}, 
Jie Li\altaffilmark{3,5},    
Robert C. Morehead\altaffilmark{11,12,13}, 
Robert L. Morris\altaffilmark{3,5}, 
Avi Shporer\altaffilmark{15,16,14,17}, 
Jeffrey C. Smith\altaffilmark{3,5}, 
Martin Still\altaffilmark{8}, 
Jeffrey Van Cleve\altaffilmark{3,5} 
}
\altaffiltext{1}{Department of Astronomy and Astrophysics, University of California, Santa Cruz, Santa Cruz, CA 95064, USA; current address: Department of Astronomy and Astrophysics, University of Chicago, 5640 S. Ellis Ave., Chicago, IL 60637, USA}
\altaffiltext{2}{Hubble Fellow}
\altaffiltext{3}{NASA Ames Research Center, Moffett Field, CA, 94035, USA}
\altaffiltext{4}{Harvard-Smithsonian Center for Astrophysics, 60 Garden Street, Cambridge, MA 02138, USA}
\altaffiltext{5}{SETI Institute, Mountain View, CA, 94043, USA}
\altaffiltext{6}{Fermilab Center for Particle Astrophysics, P.O. Box 500, MS 127, Batavia, IL 60510, USA; CIERA - Northwestern University, 2145 Sheridan Road, Evanston, IL 60208}
\altaffiltext{7}{Department of Astronomy, Box 351580, University of Washington, Seattle, WA 98195, USA}
\altaffiltext{8}{Bay Area Environmental Research Institute/NASA Ames Research Center, Moffett Field, CA 94035, USA}
\altaffiltext{9}{Department of Physics and Astronomy, San Jose State University, San Jose, CA 95192, USA}
\altaffiltext{10}{NASA Exoplanet Science Institute / Caltech, 770 South Wilson Ave., MC 100-2, Pasadena, CA 91125, USA} 
\altaffiltext{11}{Center for Exoplanets and Habitable Worlds, 525 Davey Laboratory, The Pennsylvania State University, University Park, PA, 16802, USA}
\altaffiltext{12}{Department of Astronomy and Astrophysics, The Pennsylvania State University, 525 Davey Laboratory, University Park, PA 16802, USA}
\altaffiltext{13}{Astronomy Department, University of Florida, 211 Bryant Space Sciences Center, Gainesville, FL 32111, USA}
\altaffiltext{14}{Jet Propulsion Laboratory, California Institute of Technology, 4800 Oak Grove Dr, Pasadena, CA 91109, USA}
\altaffiltext{15}{Las Cumbres Observatory Global Telescope Network, 6740 Cortona Drive, Suite 102, Santa Barbara, CA 93117, USA}
\altaffiltext{16}{Department of Physics, Broida Hall, University of California, Santa Barbara, CA 93106, USA}
\altaffiltext{17}{Sagan Fellow}

\email{fabrycky@uchicago.edu}

\begin{abstract}
We report on the orbital architectures of \Kepler\ systems having multiple planet candidates identified in the analysis of data from the first six quarters of \Kepler\ data and reported by \cite{2012Batalha}.  These data show $899$ transiting planet candidates in $365$ multiple-planet systems and provide a powerful means to study the statistical properties of planetary systems.  Using a generic mass-radius relationship, we find that only two pairs of planets in these candidate systems (out of $761$ pairs total) appear to be on Hill-unstable orbits, indicating $\sim 96\%$ of the candidate planetary systems are correctly interpreted as true systems.  We find that planet pairs show little statistical preference to be near mean-motion resonances.  We identify an asymmetry in the distribution of period ratios near first-order resonances (e.g., 2:1, 3:2), with an excess of planet pairs lying wide of resonance and relatively few lying narrow of resonance.  Finally, based upon the transit duration ratios of adjacent planets in each system, we find that the interior planet tends to have a smaller transit impact parameter than the exterior planet does.  This finding suggests that the mode of the mutual inclinations of planetary orbital planes is in the range $1.0^\circ$-$2.2^\circ$, for the packed systems of small planets probed by these observations.
\end{abstract}


\section{Introduction} 
\label{secIntro}


\Kepler data have recently revealed a windfall of planetary systems via the transit technique.
The \Kepler team is in the process of vetting candidates to rule out false positives, with a special emphasis on multiplanet candidates, which has the promise of yielding a high-fidelity ($\gtrsim 98\%$) catalog of many hundreds of planetary systems \citep{2012Lissauer}.

Previously, the \Kepler team presented planetary candidates discovered in the first four months of mission data \cite[hereafter B11]{2011Boruckib}. 
Contemporary with the B11 catalog, \cite{2011Lissauerb} (hereafter Paper I), examined the dynamics and architectures of the candidate multiplanet systems.  Paper I examined the set of period ratios, both to identify any systems that appeared to be unstable, and also to determine whether resonances played a dominant role in their formation (through trapping) or dynamics (through continued perturbation).   Another important result was that systems with many transiting planets are common, suggesting that the typical multiplicity is large \emph{and} that their orbits tend to lie in a plane to within $\sim 20^\circ$ (the fewer the typical planet number, the more coplanar the systems must be) --- see also, \cite{2011Latham}.   \cite{2012Batalha} (hereafter B13) subsequently identified candidates using the first 16 months of data.  This paper updates the above investigations of Paper I to the B13 catalog of candidates and adds two additional studies: (a) their fidelity as true planetary systems based on the apparent orbital stability of almost all of the systems and (b) mutual inclinations of planetary orbits based on their transit duration ratios. 

We begin by defining the sample of planet candidates (\S~\ref{sec_sample}), in particular how we have chosen which planet candidates to omit or update.   Next (\S~\ref{sec_closepack}) we call attention to a few closely-packed planetary pairs, and we investigate possible two- or three-planet resonances in these systems.  We discuss whether the sample of candidates obeys known orbital stability requirements (\S~\ref{sec:stab}) and implications for their purity as real planetary systems (\S~\ref{sec:fide}).  The statistical properties of the distribution of period ratios is examined in \S~\ref{sec_pratios}.  In \S~\ref{sec:durrat}, we find that the transit duration ratios in multiplanet systems limit the typical mutual inclinations to just a few degrees.  We draw comparisons to the Solar System in \S~\ref{sec:sec_sscomp}.  Finally, we restate the salient results in \S~\ref{sec:conclude}.

\vspace{0.4 in}

\section{The Sample}
\label{sec_sample}
Our sample of planet candidates is based on the KOI (\Kepler object of interest) list in the appendix by B13 (Table 9).  System numbers are denoted by the integer part, and individual planets within these systems are denoted by the decimal part, of KOI numbers.  To study these systems' dynamics, we adopt stellar masses obtained from the surface gravity (its logarithm is denoted $\log g$) and stellar radius reported by B13.  We omitted a number of candidates from this list for various reasons: (a) planets with uncertain transit periods from section 5.4 of B13; (b) those mentioned as suspect in section 1 of Paper I; (c) KOI-245.04, which has low transit S/N (11.5) and poor reduced $\chi^2=2.11$; and (d) planet candidates that are based on single transits since their periods are too uncertain for the purposes of this paper (these are denoted by negative periods in B13).  Apart from the B13 planet candidates, several groups have found additional planet candidates but we do not include these candidates here\footnote{For instance, \cite{2012Ford} found KOI-1102.03, 1102.04 and \cite{2012Fabrycky} found KOI-952.05.  The \Kepler team continues to add to the list of planet candidates and is vetting the results of the transit search through the full dataset.}.

For our analysis, we revised the stellar and planetary properties of some candidates, as follows.  We updated the period of KOI-2174.03 as described in \S~\ref{sec_closepack}.  KOI-338 had a large change in stellar radius (1~$\rightarrow$~19~$R_\odot$) from the B11 catalog to the B13 catalog, due to $\log g$ determination using a newer spectrum (12 Nov. 2010).  However, there is no signal of pulsations that generally accompanies a giant star, and the transit durations match much better with the radius of a dwarf star.  These facts suggest that either the spectroscopic result is in error, or the candidates are planets orbiting a background dwarf.  Therefore, we obtain stellar parameters for this system using the \cite{2011Brown} analysis of the photometry in the \Kepler Input Catalog, which yields $M_\star=0.96~M_\odot$ and $R_\star=1.65~R_\odot$, and adjusted the planet candidate sizes accordingly.

We also made a correction to some planet sizes due to apparently ill-conditioned fits.  The issue is that some of the B13 fits have impact parameter $b$ above 1, along with a very large value of the planet radius $R_p$.  The two conspire such that the planetary disk only skims the stellar disk, and it gives a shallow transit.  We do not believe the parameters are reliable in these cases, but that the depth is more reliable for estimating $R_p$.  For the three cases in multi-transiting candidate systems where this occurs, namely KOI-601.02, 1426.03, and 1845.02, we adopted $R_p = R_\star (\rm Depth)^{1/2}$, using stellar radius $R_\star$ and Depth reported by B13 in Table 9.  This effect was found present in B11 for KOI-961, and \cite{2012Muirhead} refined the stellar size and mass and the sizes of the planets; we adopt their parameter values in this case.

\begin{figure*}
\plotone{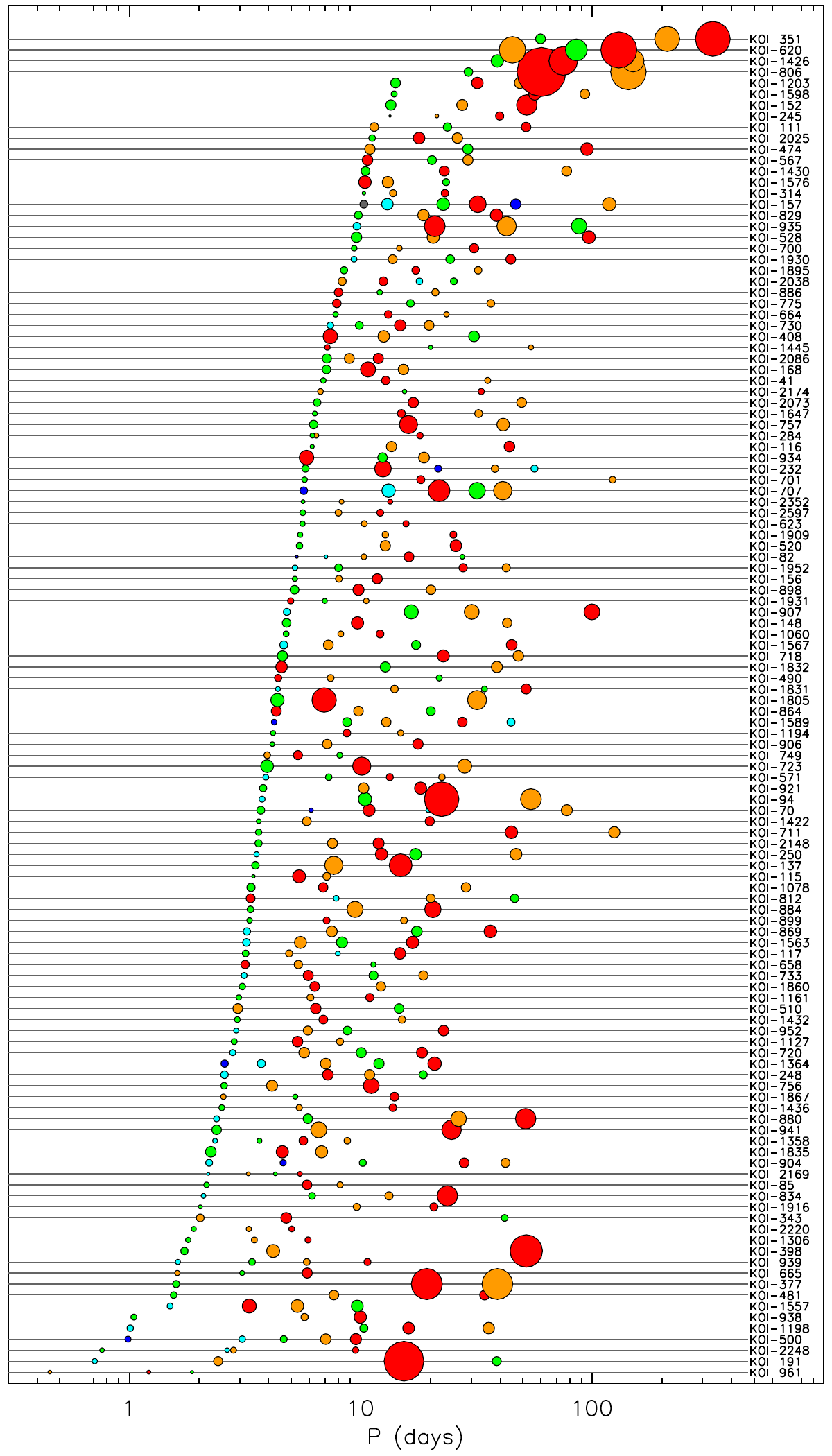} 
\caption{  Systems of three or more planets.  Each line corresponds to one system, as labelled on the right side.  Ordering is by the innermost orbital period.  Planet radii are to scale relative to one another, and are colored by decreasing size within each system: red, orange, green, light blue, dark blue, gray. \vspace{0.4 in} }
\label{fig:gallery}
\end{figure*}

With these changes from B13, the planet candidate systems are 1409 targets with a single candidate, 243 double systems, 85 triple systems, 28 quadruple systems, 8 quintuple systems, and 1 sextuple system.  This implies a total of 365 candidate multiple-planet systems with 899 candidate planets.  Parameters of these planets and their host stars are given in Table~\ref{tab:multis}, which ultimately derive from B13 (Table 9).  Overall, the number of multiple-planet systems approximately doubled from Paper I, and the largest fractional increases were seen in the quadruples ($8 \rightarrow 28$) and quintuples ($1 \rightarrow 8$).   We display the periods and sizes of planets in triple systems and above in figure~\ref{fig:gallery}.

\epsscale{0.6}
\epsscale{1.0}

\vspace{0.4 in}

\section{ Dynamics of the new systems }

In order to make inferences about the dynamical interactions of the planets, we convert their measured radii to masses according to the mass-radius relationship given in Paper I.  It is subject to the caveats that (a) the measured planetary radii, $R_p$, scale with the stellar radii, which are not always accurately known, and (b) we anticipate real planets have a diversity of structures, leading to a range of masses at any particular radius (e.g., \citealt{2011Wolfgang}).  Nevertheless, we model the systems using the simple power-law relationship for planetary masses 
\begin{equation}
M_p = M_\oplus (R_p/R_\oplus)^{\alpha}, \label{eq:mvsr}
\end{equation}
where $M_\oplus$/$R_\oplus$ are the mass/radius of the Earth, $\alpha=2.06$ for $R_p>R_\oplus$ and $\alpha=3$ for $R_p \leq R_\oplus$.  The choice of $\alpha$ for large planets, identical to the assumption we made in Paper I, is motivated by Solar System planets: It provides a good fit to Earth, Uranus, Neptune, and Saturn.  Continuing that power-law below Earth would mean smaller, rocky planets are more dense, which is not likely, whereas our choice of $\alpha = 3$ yields planets with a density equal to that of Earth.

A length scale relevant for dynamical interactions is the mutual Hill radius, given by:
\begin{equation}
R_{H} = \Big{[}\frac{M_{\rm in} + M_{\rm out}}{3 M_\star} \Big{]}^{1/3} \frac{(a_{\rm in}+a_{\rm out})}{2}, \label{eq:rhill}
\end{equation} 
where the two planets are indexed by ``in'' and ``out'', $M$ are their masses and $a$ are their semi-major axes, and $M_\star$ is the mass of the stellar host.   Relevant to stability (see \S~\ref{sec:stab} below) is the separation of their orbits in units of their Hill radii:
\begin{equation}
\Delta = (a_{\rm out}-a_{\rm in})/ R_H . 
\end{equation}
In dynamically modelling the systems, we take as initial conditions circular orbits with the periods and phases inferred from the transit observations (Table \ref{tab:multis}), with the stellar mass of B13.

The orbital period ratios (used in \S~\ref{sec_pratios}) and $\Delta$ for all 365 \Kepler multiple-candidate systems are given in Table~\ref{tab:multis}.   Some planetary systems are especially tightly packed, or lie close to three-body resonances, and we individually discuss these.  The stability properties of the ensemble of multi-transiting systems can be used to characterize the fidelity of the sample, i.e. determine whether these candidate systems have the correct periods and should be interpreted as multiple planets around the same star.  We pursue these two lines in the following. 

\subsection{Closely-spaced planets}
\label{sec_closepack}

The most closely spaced pair of new candidates are 2248.01 and 2248.04 in KOI-2248 with a period ratio of 1.065.  This pair is unlikely to be stable if both these planets are orbiting the same star (due to the separation in terms of Hill radii is likely small; see below).  The same situation is discussed in Paper 1 for KOI-284, where candidates 284.02 and 284.03 have a period ratio of 1.038.  In systems such as this, where transits are detected with a low signal-to-noise ratio, we must consider the possibility that some subset of the transits were not detected, or spurious transits were detected, thus the observed period is an alias of the true one.  We checked aliases at periods 1/4, 1/3, 1/2, 2, 3, and 4 times the nominal period by measuring the depth of the signal at locations implied by those periods.  For the KOI-2248 system, the signals are consistent with the reported periods, and inconsistent with these possible aliases.

Likely alternative explanations for this system are (a) one or both candidates is actually a blended eclipsing binary or (b) the two are true planets, but orbiting different members of a wide binary star (\citealt{2012Lissauer, 2014Lissauer}).  Following \cite{2010Steffen}, we can consider the ratio of orbital-velocity normalized transit durations:
\begin{equation}
\xi \equiv \frac{T_{\rm dur,in} / P_{\rm in}^{1/3}}{T_{\rm dur,out} / P_{\rm out}^{1/3}} ,  \label{eq:xi}
\end{equation}
where $T_{\rm dur}$ is the transit duration and $P$ is the orbital period\footnote{ When referring to pairs of planets, we use ``in'' and ``out'' to denote the inner and outer planets, respectively.}.  If this quantity is near unity, it implies that the planets are orbiting stars of roughly equal density (perhaps the {\it same} star; \citealt{2012Lissauer}).  For the unstable pairs in KOI-284 and KOI-2248, the value of $\xi$ is $0.96$ and $0.97$ respectively; this is sufficiently close to 1 that this test fails to give evidence of the pairs orbiting different stars.  However, it suggests that if the planets are orbiting different stars in a physical binary, then the two stars may be similar and resolvable with high quality imaging -- this has already been achieved for KOI-284 \citep{2012Lissauer}. 

The next closest pair is KOI-2174, with a period ratio 1.1542 between 2174.03 and 2174.01.  We performed the same alias check described above.  In this case, every other transit of the smallest planet 2174.03 (at period 7.725 days) is shallower and is marginally consistent with zero ($509\pm57$ ppm versus $105\pm63$ ppm).  Therefore we adopt an ephemeris with the period doubled (BJD = 15.4502 $\times E +$245509.8024, where $E$ is an integer), in Table~\ref{tab:multis}.

As we continue to wider period ratios, we no longer find reason on stability grounds to question the hypothesis that the systems are truly multiple planets orbiting an individual star.  We now discuss other dynamically interesting systems that are closely packed.  KOI-1665 has a period ratio 1.17219 between 1665.01 and 1665.02.   These are small candidates ($1.2$ and $1.0$~$R_\oplus$) around a solar-type star, so the alias check above is not as powerful.  Nevertheless, it raises no suspicion of the measured periods being incorrect.  Given the planets' small sizes, they are likely to have low masses and even this exteme period ratio is stable in the Hill sense (see \S~3.3, below).  KOI-262 (Kepler 50) has a nearly exact 6:5 commensurability, with a period ratio of $1.20010 \pm 0.00003$.  The planetary nature of this system was confirmed by transit timing variations \citep{2012Steffene}.  All other planet pairs have period ratios $>1.25$.  In fact, in \S~\ref{sec_pratios}, we note that there may be a significant excess of planet pairs just wide of that period ratio, with planet sizes between Earth and Neptune.  Kepler-11b and c \citep{2011Lissauera} are confirmed examples of this variety.  

\subsection{ Three-body resonances }

We also checked for potential three-body resonances among planets in systems of higher multiplicity.  Though we do not investigate whether these resonances are overabundant relative to a random distribution, we point them out because they have a dynamical effect on the systems.  Following \cite{2011Quillen}, we searched for small values of the frequency
\begin{equation}
f_{\rm 3-body} = p f_{\rm in} - (p + q) f_{\rm mid} + q f_{\rm out}, \label{eq:3bod}
\end{equation}
where $f_{\rm in}$, $f_{\rm mid}$, and $f_{\rm out}$ are the orbital frequencies (inverse periods) of the innermost, middle, and outermost planets, respectively, and $p$ and $q$ are integers.  We recovered the possible resonant chain of KOI-730, four planet candidates with period ratios near 4:3, 3:2, and 4:3, as described by Paper I.  We also found KOI-2086 (Kepler-60, \citet{2012Steffene}), whose three planets are in or near an even more closely packed chain of first-order resonances, 5:4 and 4:3, and where both neighboring pairs of planets orbiting KOI-2086 are offset by the same amount from the two-body resonances:
\begin{eqnarray}
4 f_{\rm in} - 5 f_{\rm mid} = -0.10 \pm 0.03~{\rm degrees}~{\rm day}^{-1}, \\
3 f_{\rm mid} - 4 f_{\rm out} = -0.09 \pm 0.02~{\rm degrees}~{\rm day}^{-1},
\end{eqnarray}
such that the combined three-body frequency $f_{\rm 3-body}$, with $(p,q)=(1,1)$, is $-0.004\pm0.009~{\rm degrees}~{\rm day}^{-1}$.  This is considerably closer to zero than its pair of two-body equivalents, which suggests that this resonance chain could have dynamical significance.  This fact places this system, in terms of its proximity to a multibody resonance chain, between KOI-730 and KOI-500.  In the latter, the outer four planets are more significantly offset from the two-body resonances, yet are consistent with a three-body resonance (as described in Paper I).  A final case of a possible three-body resonance is KOI-720 with planet pairs that are relatively far from two-body resonances, yet the planets $720.04$, $720.01$, and $720.03$ have $f_{\rm 3-body} = 2 f_{\rm in} - 5 f_{\rm mid} + 3 f_{\rm out}$, of $0.00 \pm 0.02~{\rm degrees}~{\rm day}^{-1}$.  This is despite another candidate planet (720.02) orbiting among them, with an orbital period greater than that of $720.01$ but less than that of $720.03$.  We  numerically integrated Newton's equations to model the four planets of KOI-720, starting them on circular orbits with the periods and phases inferred from the transit observations (Table \ref{tab:multis}), a stellar mass of $0.72 M_\odot$, and with equation~\ref{eq:mvsr} giving planet masses of $2.0$, $7.5$, $6.8$, $8.0 M_\oplus$, from the inner to outer planet.  The combination of mean motions $2 \lambda_{\rm in} - 5 \lambda_{\rm mid} + 3 \lambda_{\rm out}$ librated around $180^\circ$, with a period of $300$~years, and with an amplitude of $30^\circ$.  Thus, this three-body resonance has dynamical significance for this system, and a dedicated study of these effects seems warranted. 


\subsection{Stability of Multiple-Candidate Systems} \label{sec:stab}

Next, we investigate stability of the candidate systems.

As noted in Paper I, for two-planet systems there exists an analytic Hill-stability criterion, where the planet orbits are unable to cross (e.g., \citealt{1982MB}).    If the two planets begin on circular orbits with an orbital separation:
\begin{equation}
\Delta > 2 \sqrt{3}, \label{eq:hillcrit}
\end{equation}
then they are Hill stable \citep{Gladman:1993}.  Values of $\Delta$ are given for the observed pairs in Table~\ref{tab:multis}.  Only KOI-284 and KOI-2248 (see \S~\ref{sec_closepack}) host pairs of planet candidates that contradict this criterion.   In particular, all two-planet candidate systems obey this stability criterion, so we judge them to be plausibly stable.  

\begin{figure}
\plotone{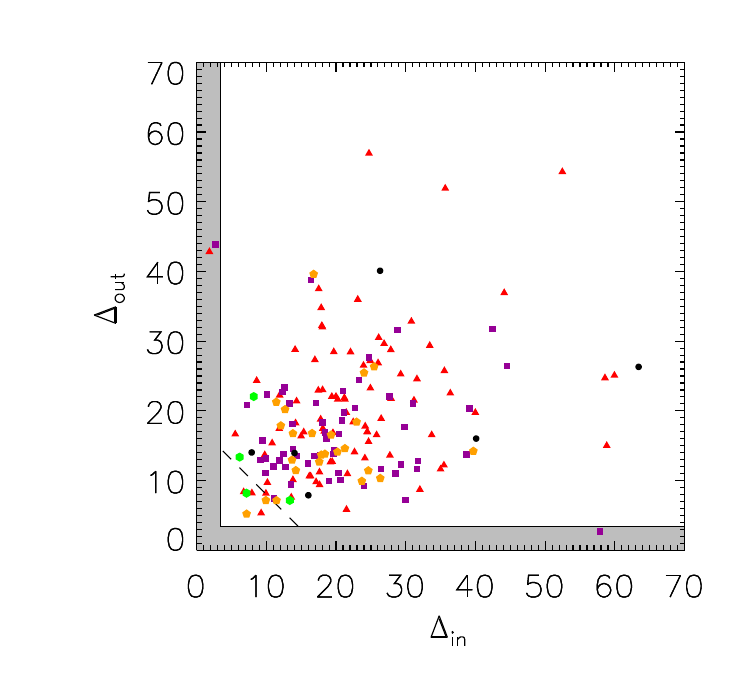} 
\caption{Separation of inner and outer pairs of triples (and adjacent 3-planet subsets of systems of three or more planets), in units of the mutual Hill separation.  The symbols denote planets in triples (red triangles), quadruples (purple squares), quintuples (orange pentagons), the sextuple (green hexagons), and the Solar System (black circles).  Systems with individual pairs that are unstable are the gray area: a triangle denoting KOI-284 and two squares denoting KOI-2248.  Other systems show three planets with particularly close spacing (below the dashed line, which is criterion \ref{eq:3plstab}), but these were numerically integrated and found to be long-term stable. \vspace{0.4 in} }
\label{fig:dd}
\end{figure}

We are aware of no analytic stability criterion for the systems with more than two planets.  However, in systems of three or more planets, the instability time scale generally increases with separation, as in the two-planet case \citep{1996Chambers, 2009Smith}. In Paper I, we numerically integrated all of the systems with more than two planets for $10^{10}$ orbits of the innermost planet using \emph{MERCURY} \citep{1999Chambers}.  We started from circular, coplanar orbits and used our power-law mass-radius relationship.  In addition to each pair obeying two-planet stability criteria, we suggested a conservative heuristic criterion,
\begin{equation}
\Delta_{\rm in} + \Delta_{\rm out}>18, \label{eq:3plstab}
\end{equation}
where the ``in'' and ``out'' subscripts pertain to the inner pair and the outer pair of three adjacent planets.  This latter criterion does not assure stability (particularly if planets are eccentric), though it suggests there is no reason to suspect the system would be unstable, based on the planet sizes and periods sensed by transits.  In Figure~\ref{fig:dd}, we plot the $\Delta$s for inner and outer pairs of threesomes.  Systems satisfying criteria~\ref{eq:hillcrit} and \ref{eq:3plstab} we do not analyze further, but the other systems may be unstable and call for further analysis.  Most of these systems were already examined in either Paper I or \citet{2012Lissauer}.  We numerically integrated the remaining three,  KOI-620 (Kepler-51, \citet{2012Steffene}), KOI-1557, and KOI-2086 (Kepler-60), as described in Paper I.  We found them to be plausibly stable.  That is, starting on circular, coplanar orbits matching the phase and periods of the data, they suffered neither ejection, nor collision, nor a close encounter within 3 mutual Hill radii over a timespan of $10^{10}$ orbits of the innermost planet (usually of order $10^8$ years).  We also integrated KOI-961 for the same duration using the masses of \cite{2012Muirhead}, and found them to be similarly stable.

\epsscale{1.0}
\epsscale{1.0}

The only new system that is unstable was KOI-2248, discussed in \S\ref{sec_closepack}.  Using the Burlisch-Stoer integrator in \emph{MERCURY}, the planets began violent gravitational scattering in several synodic time scales.  Clearly, this system needs a qualitatively different understanding for its architecture, as noted above.  One final system where at least one new planet appears close to instability is KOI-707 = Kepler-33.  An analysis of the stability of this system was carried out in the discovery paper \citep{2012Lissauer}, so we performed no additional analysis here. 

These outcomes of our stability analysis are for the power-law $M_p$--$R_p$ relationship (eq.~\ref{eq:mvsr}) with $\alpha=2.06$.  To see how many systems would be unstable if the planets were denser, we considered a larger $\alpha$ value for planets below $2 R_\oplus$ (an estimate of the Super-Earth / mini-Neptune boundary).   We looked for $\Delta < 2 \sqrt{3}$ for any adjacent pair.  Below $2 R_\oplus$, for any $\alpha$ below $6.9$, no additional systems violate Hill's stability given circular orbits.  Therefore all these planets may have an Earthlike composition, for which $\alpha \simeq 3.7$ \citep{2006Valencia}, and not violate stability limits.

For planets larger than $2 R_\oplus$, we may also consider denser planetary structures, by varying $\alpha$.  No additional systems display instability for $\alpha \leq 2.6$ [i.e., $M_p = M_\oplus (R_p/R_\oplus)^{2.6}$].  For the slightly higher value of $\alpha=2.7$, the pair of planets of KOI-523 and the outer two planets of KOI-620 would be unstable, according to our numerical integrations.  In KOI-523, the planets would have (mass, radius) of ($0.99 M_{\rm Nep}$, $0.72 R_{\rm Nep}$) and ($1.99 M_{\rm Sat}$, $0.74 R_{\rm Sat}$), i.e. a Neptune-mass planet that is $2.6$ times as dense as Neptune, and a planet twice as massive as Saturn but that occupies less than half of Saturn's volume.  In KOI-620, the planets would have (mass, radius) of ($1.16 M_{\rm Sat}$, $0.60 R_{\rm Sat}$) and ($1.44 M_{\rm Jup}$, $0.86 R_\oplus$): more massive, yet much smaller versions of the Solar System's gas giants.  Such a large $\alpha$ would imply an extreme density for gas-giant planets, even exceeding that of the core-heavy transiting planet HD 149026 \citep{2005Sato}.  From this exercise, we see that our conclusions about stability are not sensitive to our adopted masses.  Conversely, stability considerations give us little useful constraint on these planets' physical structure. 

To summarize this stability study, for all the pairs of planet candidates, only two are suspected to be unstable on million-year timescales given low eccentricities and inclinations: KOI-284 and KOI-2248.  Higher multiplicities do not appear unstable either, based on numerical integrations.  Using a mass-radius relationship favoring high density, a few more systems could be unstable.  We repeat the caveat that we have only considered instability while using initially planar, circular orbits; if these systems contain planets in eccentric orbits, they would likely be less stable.

\subsection{Fidelity of Multiple-Candidate Systems} \label{sec:fide}

\cite{2011Morton} have emphasized that planet candidates from \Keplert, once properly vetted using indicators in the data itself, tend to be highly reliable ($> 90$\%), and \cite{2012Lissauer} extended and strengthened this statement for candidate multiple-planet systems.  The density of background eclipsing binaries is so low, and the small depth and detailed shape of transits is unlikely to be mimicked because of the photometric precision of \Keplert, that more than one pattern of transit signals on a single target are unlikely to occur via a blending of stellar eclipses with additional stellar light.  Moreover, \Keplers photometric precision and centroid analyses means transit events occurring on background stars must lie very near the target star, in projection, which is unlikely.

We can now address the statistical reliability of \Keplers multiplanet candidates from a new and independent angle.  With so few candidate planetary systems showing instability (2 out of 761 pairs, including the higher-order multiples), we expect most of these candidate systems are real planetary systems.  Consider the possibility that pairs are ``split multis,'' defined as a system that appears to be a pair of planets around a star, but the events are actually split into more than one system.  The most likely alternatives are (a) one or both members of the pair of planetary candidates is a blended eclipsing binary, or (b) both planet candidates are planets, but they orbit different stars \citep{2012Lissauer}.  Such cases need not obey stability constraints.  Therefore we can estimate an expected fraction of apparently unstable systems, given the hypothesis that all these candidate systems are split multis.

If we draw two planets from the ($P$, $M_p/M_\star$) values of all the planet candidates in multis, and consider whether that pair would be stable if in the same system, $\Delta \leq 2 \sqrt{3}$ occurs in $25,867 / 403,651 \simeq 6.4\%$ of the draws (the integer numbers are computed from sampling all the possible pairs).  That is, one would expect $48.8$ pairs to be unstable over the whole set of $761$ pairs.  Using the Poisson distribution, to have found two or fewer unstable pairs given the expectation value $\lambda=48.8$ has a tiny probability of $8 \times 10^{-19}$.

\begin{figure}
\plotone{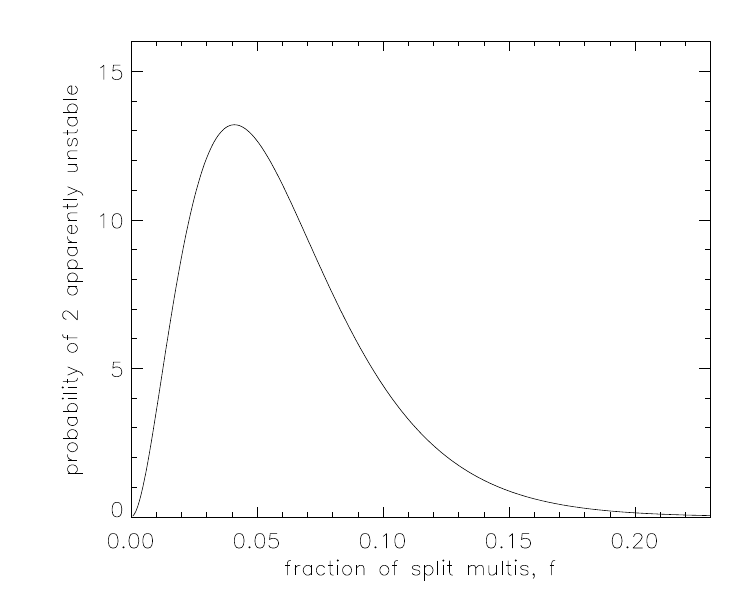} 
\caption{ Normalized probability distribution of the fraction of ``split multis,'' $f$.  Assuming a uniform prior probability on $f$, this posterior probability function is derived by Bayes' theorem (equation~\ref{eqn:bayes}), conditioned on the ``data'' that two observed pairs are Hill-unstable out of 761 possible pairs.  \vspace{0.4 in}}
\label{fig:frac}
\end{figure}

On the other hand, if only a fraction $f$ of the systems are split multis, then the expected value of apparently unstable systems falls to $\lambda f$.  Given that only two systems in our sample appear to be unstable, we can place Bayesian constraints on the fraction $f$.  Let us take a prior probability distribution of $f$ which is uniform from 0 to 1: $p(f)=1$.  Then we can apply Bayes' theorem to estimate the probability of $f$ given the observations:
\begin{equation}
P(f | {\rm data}) = \frac{ P( {\rm data} | f ) } { \int_0^1 df' P( {\rm data} | f')}, \label{eqn:bayes}
\end{equation}
where $P( {\rm data} | f )$ is the probability of the data given $f$ and $f'$ is an integration variable used to determine the normalization.  The only information we use (i.e., the ``data'') is that there are two apparently unstable systems.  This probability distribution is given in figure~\ref{fig:frac}, which shows a mode of 4.1\% and a wide range of possible fractions: the 95\% credible interval is 1.3\% -- 14.7\%.  These estimate are marginally larger than the $\lesssim 2\%$ of the candidates in multiple systems not being true planets estimated by \cite{2012Lissauer}. 
We note that for the present estimate, we are counting planets that are around two different stars in a physically bound binary as a split multi, which as discussed in \S~\ref{sec_closepack}, is likely to account for one of our unstable pairs, KOI-284, and may well account for the other, KOI-2248.

\epsscale{0.9}
\epsscale{1.0}

These estimates are based on drawing an ensemble of $P$ and $M_p/M_\star$ values, which were in turn assumed to follow certain distributions, so let us examine the robustness of the conclusions to varying those assumptions.  First, we chose a period distribution $P$ matching the planet candidates in multiple systems.  This distribution nearly matches the single-candidate period distribution, so this is appropriate if the split multi hypothesis is that a pair of planets are actually singles around hosts that are blended together.  However, this distribution is narrower than the detached eclipsing binary distribution, which may be blended into some of the targets to produce the split-multi signal.  To explore this, we selected $M_p/M_\star$ as above (a reflection of the distribution of observed depths) but replaced the periods by two draws from the list of eclipsing binaries labeled ``detached'' by \cite{2011Slawson}.  This was done in a Monte Carlo fashion, resulting in an unstable fraction $\lambda / 761 = 5.07\% \pm 0.04\%$.  Given that this expectation is lower than above, the fraction of split multis would need to be higher in order to produce two apparently unstable systems: assuming the periods are drawn from the eclipsing binary distribution, the 95\% credible interval of $f$ spans $1.7$\% -- $18.6$\%.  The second assumption is the particular mass-radius relationship we adopted, which gave $M_p/M_\star$.  If the planets are actually denser than assumed, more systems would be deemed unstable.  Above we tested the sensitivity of our stability results to varying the mass-radius relationship for the known systems, and we found that extreme densities are needed for any additional planetary systems to be unstable.  For these random pairs, the number of unstable systems expected would vary smoothly with their assumed masses, and hence the range of $f$ would vary as well. 

By considering stability, we have seen that $\sim 96\%$ of the pairs of multi-transiting candidates are actually planets around the same star.  Recall that this estimate is independent of that by \cite{2012Lissauer}, who used binary statistics to estimate that in fully vetted systems, $\gtrsim 98\%$ are real planets.  In the following sections we rely on such high fidelity, assuming that all the systems are real as we characterize their architectures.  Because of their apparent instability, from this point on we cull KOI-284.02, KOI-284.03, KOI-2248.01, and KOI-2248.04.  KOI-284 becomes a single-planet system and is not included in the analysis, and KOI-2248 becomes a two-planet system and is analyzed as such.


\section{Period ratio statistics}
\label{sec_pratios}

\begin{figure*}
\plotone{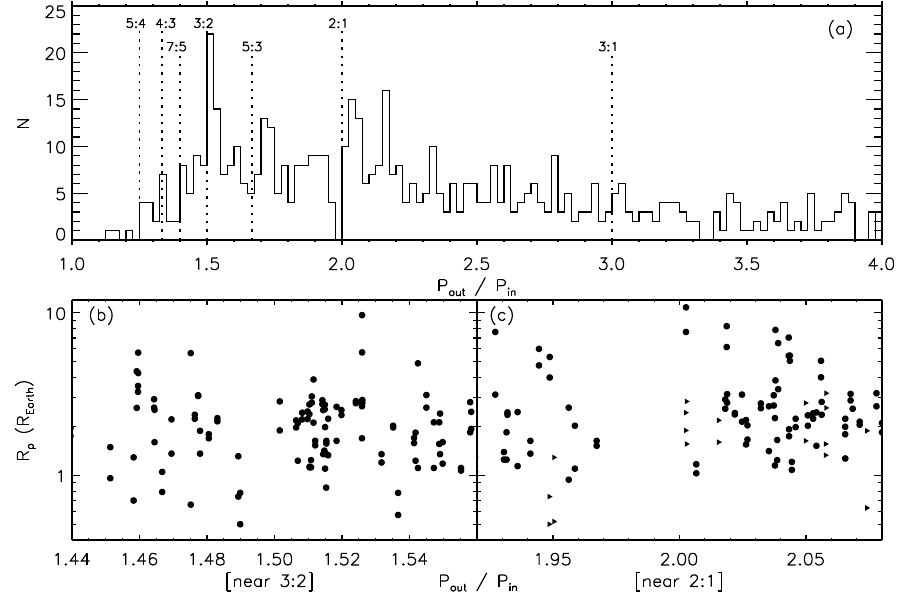} 
\caption{ Period ratio statistics of all planet pairs.  Panel (a): Histogram of all period ratios in the sample (i.e., pairwise between all planets in higher order multiples, not just adjacent planets), out to a period ratio of 4.  First order (top row) and second order (lower row) resonances are marked.  The mode of the full distribution is slightly wide of the 3:2 resonance, and there is an asymmetric feature near the 2:1 resonance.  There are few systems interior to the 5:4 resonance.  Panel (b): Planetary radii versus the period ratio for planetary pairs near ($0.04 \pp$) the 3:2 resonance.  Both radii for each pair are plotted.  Panel (c): same as panel b, but near (within $0.04 \pp$) the 2:1 resonance.  Triangles denote planet pairs that are not adjacent, or in other words have an intervening transiting planet. \vspace{0.4 in}}
\label{fig:pratboth}
\end{figure*}

In figure \ref{fig:pratboth} we plot a histogram of the period ratios ($\pp \equiv P_{\rm out}/P_{\rm in}$) of all pairs, not just adjacent pairs, in all systems.  It spans a wide range, from hierarchical configurations to the edge of stability.  There is an apparent cut-off interior to the 5:4 resonance, however KOI-262 (Kepler 50, \cite{2012Steffene}) is a known system near 6:5 and KOI-277 (Kepler 36, \cite{2012Carter}) is near 7:6 (though it was not included in this study since the smaller planet was not identified in B13).  The existence of these two systems show that this region interior to the 5:4 resonance is not empty.   As can be seen in figure \ref{fig:pratboth}, the main conclusion of Paper I remains: The vast majority of planet pairs are not in resonance.  However, as resonances do have dynamical significance, we address their statistical properties in this section. 

\epsscale{1.0}

\begin{figure}
\plotone{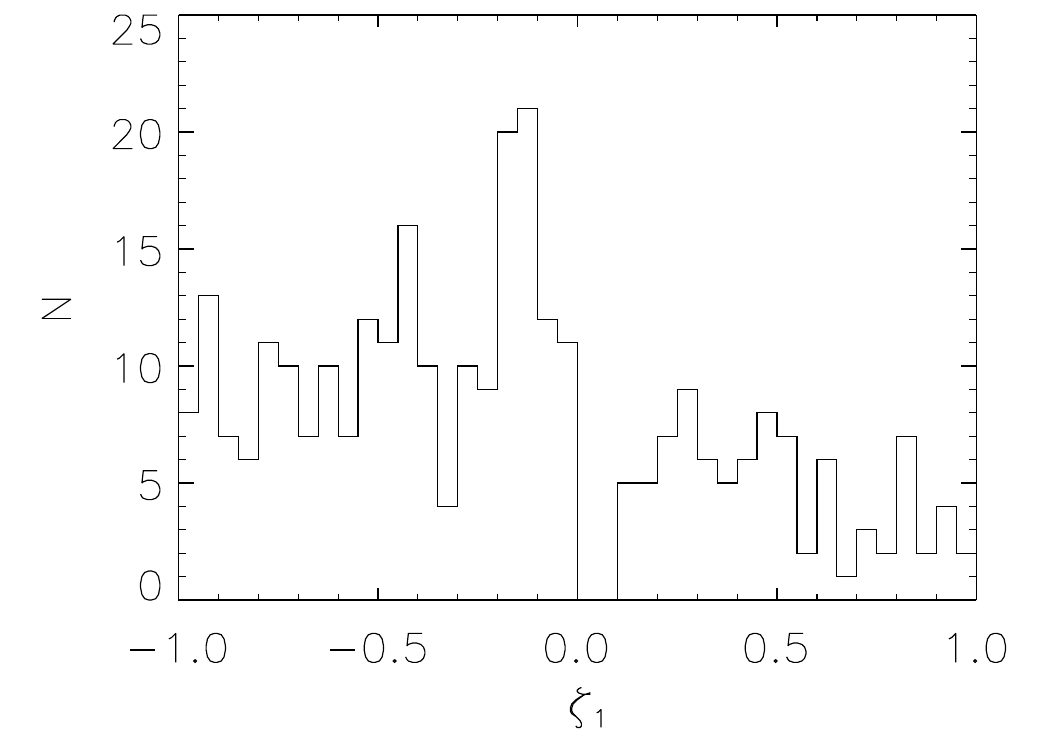} 
\caption{ Histogram of $\zeta_1$, a variable describing the offset from first-order resonances (eq.~\ref{eq:zeta1}), for all planetary pairs in the neighborhood of a first-order resonance, i.e. with a period ratio between $1$ and $2.5$.  The spike between $-0.1$ and $-0.2$ means that period ratios just wide of first-order resonances are overpopulated relative to a random and even distribution.\vspace{0.4 in} }
\label{fig:zeta}
\end{figure}

To study the properties of first-order resonances, we compute the $\zeta_{1}$ variable introduced in Paper I:
\begin{equation}
\zeta_{1} = 3 \left( \frac{1}{\pp - 1} - {\rm Round}\left[ \frac{1}{\pp - 1 } \right] \right), \label{eq:zeta1}
\end{equation}
which describes the distance a pair of planets is from a first-order resonance.  The variable $\zeta_{1}$ has a value 0.0 when the period ratio is $\pp= (j+1)$:$j$, i.e., first-order resonances and its ``neighborhood'' extends to $-1$ and $+1$ at the adjacent third-order resonances interior and exterior to the first-order resonance, or at $(3j+2)$:$(3j-1)$ and $(3j+4)$:$(3j+1)$, respectively.  In figure \ref{fig:zeta} we plot the histogram of $\zeta_1$, with all values of $j$ and all planetary pairs contributing.  As in Paper I, we find an excess of planet pairs with $-0.2 < \zeta_{1} < -0.1$, i.e., \emph{pairs of planets prefer to be just wide of first-order resonances}.

We compare the observed $|\zeta_1|$ distribution to a random distribution, which is uniform in the logarithm of period ratios, via a Kolmogorov-Smirnov (K-S) test.  The null hypothesis is that period ratios are smoothly distributed, e.g. that they do not occur more often near ratios of integers (which correspond to dynamical resonances).  A significant difference in these distributions is detected with $p$-value $=1.4\times10^{-3}$, which implies that the distribution is peaked within a few percent of the first-order resonances.  With the addition of the new systems, this number is little-changed from that of Paper I (where it was $1.2\times10^{-3}$).  In Paper I it was found that a different variable, $\zeta_2$, hinted that second-order resonances might be distributed similarly ($p$-value $0.046$).  However, we find with this expanded sample that $|\zeta_2|$ is now more consistent with a logarithmically-uniform distribution of period ratios, with K-S test $p$-value = $0.082$.  Nevertheless, some specific systems (e.g., KOI-738 = Kepler-29, \citealt{2012Fabrycky}) are in or near dynamical second-order resonance.  We describe a more general formalism for the $\zeta$ variable in appendix~\ref{sec:zz}, which gives context to our choice of equation~(\ref{eq:zeta1}) and may be useful in future investigations of the statistics of resonance. 

Let us explore this preference for first-order resonances further.  First, we compare the observed $|\zeta_1|$ distribution to a random distribution in the neighborhood of 2:1 (between 7:4 and 5:2).  The distributions do differ, with a $p$-value of $0.031$; however this has weakened from $0.00099$ (in Paper I) with the expanded sample considered here which includes more small planets.  The more important effect contributing to the first-order resonance result is that systems in the neighborhood of 3:2 (between 10:7 and 8:5) tend to be near 3:2; $|\zeta_1|$ differ from a random distribution in this neighborhood with a $p$-value of $0.0071$.  Looking at panel (a) of figure~\ref{fig:pratboth}, the global peak is just wide of the 3:2 resonance; a comparatively smaller peak exists just wide of 2:1.  The peak at 3:2 appears to be a true excess of systems where the integrated population near the 3:2 remains above the baseline.  On the other hand, the peak just wide of 2:1 contains only slightly more pairs than the trough just narrow of the 2:1 is missing, possibly indicating that near the 2:1 resonance planetary orbits are ``redistributed'' from where they nominally formed.  

For a better view of these resonances, we plot the period ratios of individual planet pairs near $1.5$ and $2.0$ in panels (b) and (c), respectively, of figure \ref{fig:pratboth}.   Just wide of $1.5$, we note a dense cluster (spanning $1.505$ to $1.520$ for $R_p \lesssim 3.0~R_\oplus$).  A similar over-density wide of  $2.0$ is apparent, but it is considerably more diffuse.  These are the main features that imply $|\zeta_{1}|$ is not evenly distributed.  In these panels, we see more clearly the lack of pairs just narrow of the resonances, particularly for the 2:1 resonance.  In both cases, this gap may be slightly wider at larger planet sizes.  Insofar as planet masses correlate with planet radii, this feature may result from resonances being wider for more massive planets.  To actually generate these gaps in the period ratio distribution, additional forces need to be invoked.  These may simply be gravitational scattering, as in the case of the Kirkwood gaps in the asteroid belt, where a resonance chaotically pumps up the eccentricity \citep{1983Wisdom}, and the body scatters off other planets and is removed from the resonance.  Chaos was also noted by \cite{1986Murray} in the 3:2 and 2:1 resonances at low eccentricity, which might be sufficient to produce the gaps in figure~\ref{fig:pratboth}, panels (b) and (c).  Another possibility is the action of tidal dissipation in the inner planet, pulling it towards the star and increasing the period ratio \citep{2003Novak,2007Terquem}.  Yet another possibility is that, while the pair is still embedded in a gaseous disk, one planet may excite density waves at its resonance location that interact with the other planet, preventing resonance capture \citep{2012Podlewska}.
 
Last, we consider whether the pairs of planets near first-order resonances are statistically closer to resonance than would be expected with random spacings.  For instance there are KOI-730 (4:3, 3:2, 4:3), KOI-2086 (5:4, 4:3), and KOI-262 (6:5), which we have already discussed.   In addition, there is KOI-1426.02/1426.03, which are gas giants near the 2:1 resonance.  All these cases lie in the region $|\zeta_{1}|<0.05$, however, they do not cluster near $\zeta_{1} \simeq 0$ significantly more than random.  Thus, while these pairs appear to be unusually close to exact resonances ($\delta {\cal P} / {\cal P} < 0.001$) and their dynamics is likely dominated by those resonances, they may simply be members of the smooth distribution of period ratios.  If true, this would indicate that they are not necessarily the product of differential migration that would produce an excess population near resonance.

In systems with multiple, adjacent first-order resonances, the candidates are more likely to be bone fide planets (Paper I).  Taking as the null hypothesis a uniform spacing in $\log {\cal P}$ (i.e., that near-resonant locations are not preferred), the distribution of $\zeta_{1}$ is nearly uniform, indicting that two adjacent period ratios have $|\zeta_{1,{\rm in}}| + |\zeta_{1,{\rm out}}|$ less than or equal to a small value $x$ with a probability $p \simeq x^2/2$.   (This is actually conservative estimate, as a logarithmic distribution of $\log {\cal P}$ yields a slightly lower probability than a uniform distribution at small $\zeta_{1}$.)  For the case of KOI-2086 (Kepler 60), the values of the two adjacent spacings are $\zeta_{1,{\rm in}}=-0.0324$ and $\zeta_{1,{\rm out}}=-0.0276$.  Thus, such systems would be this close to a first-order resonant chain only $p=0.18$\% of the time.

Given $n=169$ sets of three adjacent planets, the expectation value that at least one of them will show such a close chain is $1-(1-p)^n = 26$\%.  Therefore, KOI-2086's chain is not unexpected even if planetary pairs do not prefer resonances.  Having 4 planets in a resonant chain would be less expected, and having $|\zeta_{1,{\rm in}}| + |\zeta_{1,{\rm mid}}|+ |\zeta_{1,{\rm out}}|$ (where subscript `mid' refers to the middle pair) less than or equal to $x$ occurs with probability $p \simeq x^3/6$.  For KOI-730, $\zeta_{1,{\rm in}}=-0.0123$, $\zeta_{1,{\rm mid}}=-0.0186$, $\zeta_{1,{\rm out}}=-0.0063$, and thus $p=8.6\times 10^{-6}$.  There are $n=47$ sets of 4 adjacent planets, so the expectation value that at least one would show such a chain is $1-(1-p)^n = 0.04$\%; i.e., a multi-resonant chain like that in KOI-730 is very unlikely if the orbital periods of planet candidates with a common host star were completely independent.

\section{Duration ratio statistics and coplanarity}
\label{sec:durrat}

The durations of planetary transits were recognized to be a source for information on orbital eccentricity well before the \Kepler launch, as the eccentricity causes the orbital speed to differ from the circular case, and transit duration is inversely proportional to projected orbital speed \citep{2008Ford}.  Using the B11 KOI catalog, \cite{2011Moorhead} performed an analysis of the statistics of durations and found evidence for moderate eccentricities among small planets.  This result required knowledge of the stellar masses and radii.  Several authors \citep{2010Ragozzine, 2011Kipping} have also pointed out that the properties of the star (most directly, its density) can be constrained by the durations and ingress and egress time of the transits, especially in systems with multiple planets.  In such cases no detailed stellar model is needed, and constraints on the eccentricities of the planets are by-products.  Finally, it has been noted that the relative transit durations of planets present in the same lightcurve can be used to validate them as planets around the same host star (\citealt{2011Morehead, 2012Lissauer}).

\subsection{Duration ratios: method} \label{sec:durmeth}

Here we \emph{assume} the planetary candidates are in true systems orbiting the same star, and we investigate how the distribution of duration ratios depends on coplanarity.  In the limit that all the planetary orbits within a system are circular and coplanar, the impact parameters $b$ and semi-major axes $a$ have the relationship:
\begin{equation}
b_{\rm out} / b_{\rm in} = a_{\rm out} / a_{\rm in} \quad  [\emph{coplanar, circular}] , \label{eqn:brat}
\end{equation}
where ``in'' signifies an interior planet and ``out'' signifies an outer planet.  Thus we expect that $b_{\rm out}$ will be larger than $b_{\rm in}$ in systems where both planets are close to coplanar and both transit.  At the other extreme, planets around the same star may be sufficiently misaligned to destroy such correlations, which requires a typical mutual inclination $i \gtrsim R_\star / a$, where $R_\star$ is the host's radius and $a$ is a typical semi-major axis.  In that case, both $b_{\rm in}$ and $b_{\rm out}$ would be drawn from the same distribution, and $b_{\rm in}$ would be larger than $b_{\rm out}$ half the time.  Such a cases have been observed, for instance: in the Kepler-11 e/g and Kepler-10 b/c pairs, the observed $b_{\rm out}$ is smaller than that given by equation~(\ref{eqn:brat}), and the orbits must deviate from coplanarity by at least $1^\circ$ and $5^\circ$, respectively, for these particular pairs.  Thus we expect the distribution of impact parameters can help us determine the distribution of mutual inclinations.

We do not have sufficiently accurate stellar properties or good knowledge of impact parameters themselves to perform such comparisons directly.  However, transit durations $T_{\rm dur}$, from first to fourth contact, are generally well-measured and are $\simeq 2 ((1+r)^2-b^2)^{1/2} R_\star / v_{\rm orb}$, where $r \equiv R_p/R_\star$ and $v_{\rm orb} \propto P^{-1/3}$.  Therefore for each planet in a system, the function $((1+r)^2-b^2)^{1/2}$ is proportional to $T_{\rm dur}/P^{1/3}$.  The ratio of this latter quantity for the pair of planets, $\xi$ (eq.~\ref{eq:xi}), is the quantity that is precisely measured and is sensitive to the mutual inclination of the orbits through their relative impact parameters.  Given the distribution of $b_{\rm in}$ of inner planets will be biased towards smaller values if the systems are nearly coplanar, and in most cases $r \ll 1$, we expect the $\xi$ to be \emph{greater} than 1 for nearly coplanar systems.  In the limit that misalignment removes impact-parameter correlations, $\xi$ and $\xi^{-1}$ will have the same distribution. 

To simulate the observed $\xi$ distribution, we should take into account photometric noise and eccentricities.  Photometric noise typically introduces a relative uncertainty of $\sim1$\% in a duration measurement: $\sigma_{\rm dur} \simeq T_{\rm dur} (2 r)^{1/2} / {\rm S/N}$ \citep{2008Carter}.  We add a gaussian-random deviate with this standard deviation to the simulated durations.  Eccentricity has two effects on the duration: (i) at a given inclination, due to a different star-planet separation, it results in a different impact parameter and transit chord and (ii) the projected orbital speed differs from a circular orbit, as does the speed projected on the sky plane; we model both of these effects with Keplerian orbits with uniform-random periastron angle $\omega$.  With these effects in place, the population model assumes that mutual inclinations of planetary orbits are excited to a scale $\delta$, and eccentricities of both planets are excited to a scale a factor $n$ times $\delta$.  That is, the energy in the eccentricity epicycles is a certain number times the energy in the inclination epicycles.

Both the mutual inclination and eccentricity distributions are modeled as Rayleigh distributions, such that the Rayleigh widths are $\sigma_i = \delta$ (in radians) and $\sigma_e = n \delta$.  This allows us to use a Monte Carlo method to create simulated distributions of $\xi$ as a function of $\delta$ and $n$.   To evaluate this distribution, we make 250 mock transit systems for each observed pair of planets, where we have taken only the pairs where both planets are detected at S/N$>7.1$,  the nominal detection limit.

For each mock system we first draw the eccentricities $e \sin \omega$ and $e \cos \omega$ from Gaussian distributions of width $n \delta$ (resulting in a uniform distribution of $\omega$ values and a Rayleigh distribution of $e$ values).  We discard a trial if either planet's eccentricity is above 1 or if the inner planet's apocenter distance exceeds the outer planet's pericenter distance, given their period ratio --- a simple stability criterion.  Step two is to draw $b_{\rm out}$ uniformly within [0, $b_{\rm out,max}$], where $b_{\rm out,max}$ is the impact parameter the planet would need for the total S/N of the outer planet to drop to 7.1.  This modeled S/N is taken as the observed S/N times the square root of the ratio between the modeled duration and the observed one.  Step three is to draw $b_{\rm in}$ from a distribution centered on $b_{\rm out} (a_{\rm in}/a_{\rm out})$ (eq.~[\ref{eqn:brat}]) and having a gaussian $\sigma=\delta a/(R_\star+R_p)$ (resulting in a Rayleigh distribution of width $\sigma_i=\delta$ in mutual inclination).  If $|b_{\rm in}|>b_{\rm in,max}$, as above, this planet would not be detected in transit.  If the conditions for acceptance are not met at each of these three steps, the process begins anew at step one.  If accepted, the mock system's $\xi$ value contributes to the simulated $\xi$ distribution.  We compare these models for the distribution of $\xi$ to the data with statistical tests. 

\subsection{Duration ratios: results}

\begin{figure}
\plotone{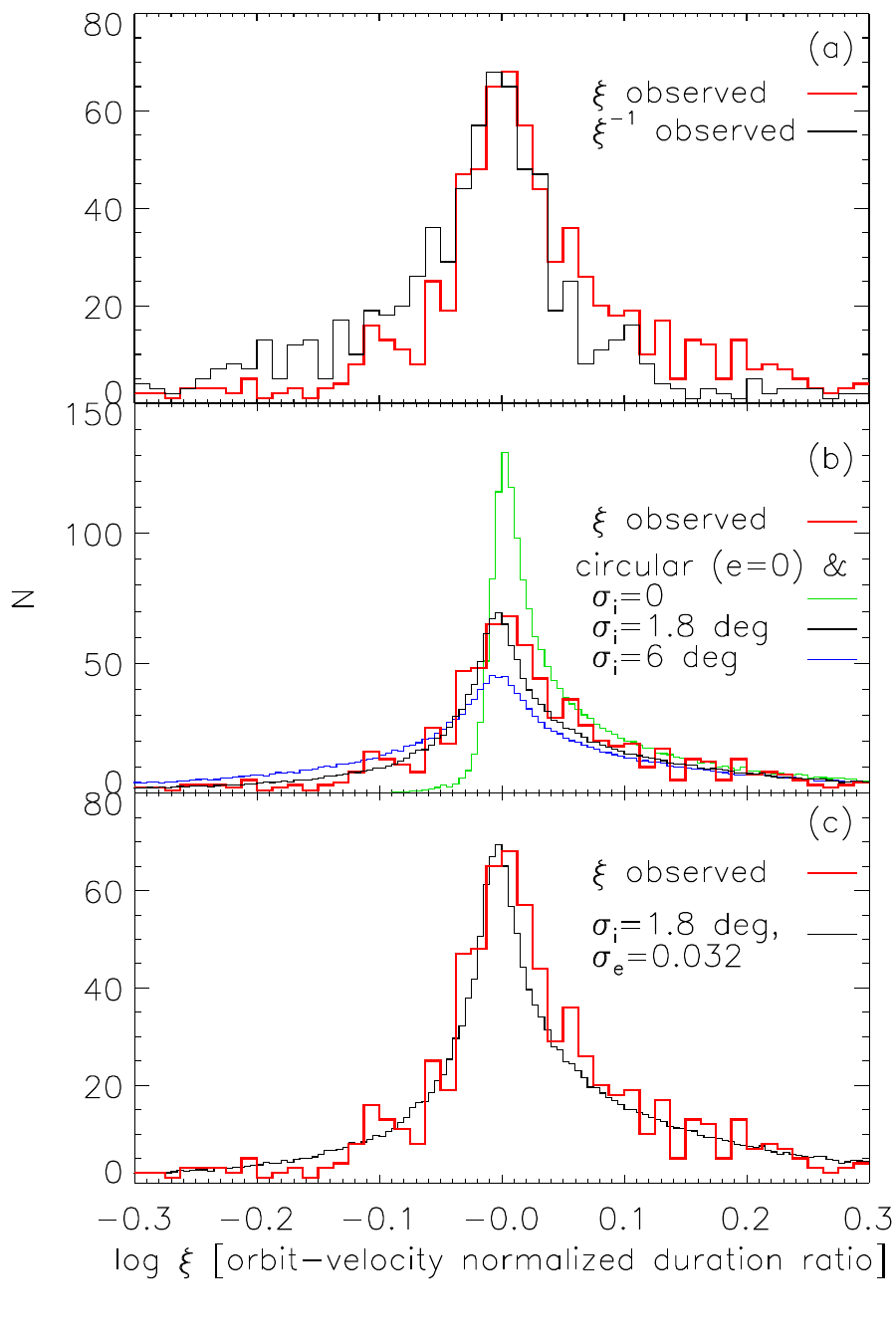} 
\caption{ Histograms of normalized duration ratios (equation~\ref{eq:xi}).  Panel (a): the distributions of the observed $\xi$ and its inverse are contrasted.  If planetary orbital planes are not correlated with each other, these distributions would be equal.  Instead, the difference in the histograms implies the inner planets have a longer duration, i.e. a smaller impact parameter, on average.  Panel (b): models of three different typical mutual inclinations, for circular orbits, are compared to the data, showing how these can be distinguished.  Panel (c): the best-fitting model is compared to the data.  This fit is not significantly better than the black line of panel b, as a wide range of eccentricities acceptably fits the data.   \vspace{0.4 in} }
\label{fig:durdiff}
\end{figure}

Let us first test the null hypothesis that planets around the same star are sufficiently misaligned to destroy the normalized-duration ratio signature discussed above.  Assuming their impact parameters are drawn from the same distribution, $T_{\rm dur,in}/P_{\rm in}^{1/3}$ and $T_{\rm dur,out}/P_{\rm out}^{1/3}$ would be distributed in the same manner, therefore their ratios $\xi$ and $\xi^{-1}$ should also be from the same distribution.  We test that in figure \ref{fig:durdiff}, panel (a), where the null hypothesis is that the black and red histograms are equivalent.  These histograms are not equal, with the center-of-mass of $\xi$ lying at a significantly larger value than $\xi^{-1}$, a K-S $p$-value of $5\times10^{-15}$.  This is the signature of planetary orbits lying in nearly the same plane. 

\epsscale{0.9}
\epsscale{1.0}

There are potential biases that could affect this conclusion.  First, the outer planet's radius is typically larger than the inner one (perhaps due to detection limits; see Paper I), but this would bias $\xi$ to values slightly less than 1, and we observe the opposite.  Another aspect is that the box-least squares search that found most of these candidates (B13) was run over a range of durations $0.003P$ to $0.05 P$.  Planets outside this range were still found, but not with an optimal matched filter.  The search algorithm is less sensitive to the very shortest durations (largest impact parameters) at long period and the very longest durations (smallest impact parameters) at short period.  Therefore this effect should bias $\xi$ downwards, again against the observed trend.  We have not identified other instrumental or analysis biases that could push the distribution to $\xi>1$ values, as observed. 

Although this test shows the observed $\xi$ distribution is asymmetric, it is indeed quite broad.  An ideal model distribution consisting of perfectly coplanar and circular systems would lie entirely above 1, due to the relation in equation~\ref{eqn:brat}.  Measurement error introduces additional spread at the few-percent level; modeling this effect gives the green curve in Figure~\ref{fig:durdiff}, panel (b), which departs only slightly from this ideal.  

\begin{figure}
\plotone{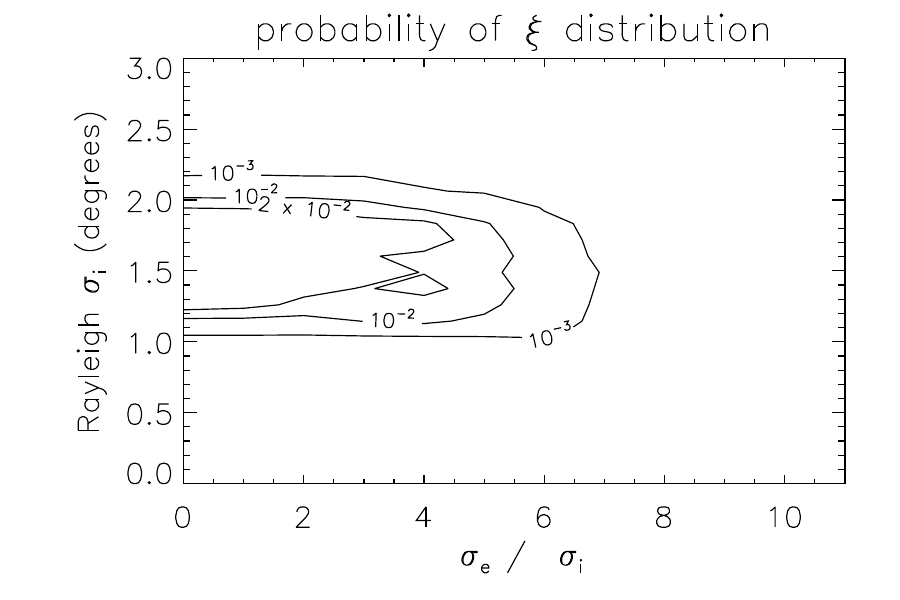} 
\caption{ Kolmogorov-Smirnov $p$-value for inclined and eccentric systems.  A region of acceptable probability lies in the range $\sim 1.0^\circ - 2.2^\circ$, for the dispersion of the Rayleigh distribution of the mutual inclination.  The acceptable range of eccentricity is large, from perfectly circular to several times equipartition with the mutual inclination.  \vspace{0.4 in} }
\label{fig:probei}
\end{figure}

Therefore, some mutual inclination or eccentricity is clearly needed.  To model these, we computed a grid of models (described in \S~\ref{sec:durmeth}) with steps of $0.002$ in $\delta$ and $1$ in $n$, and we show in figure~\ref{fig:probei} the $p$-value from the K-S test for these models.  The peak (best-fit) value has a probability $0.033$ and lies at $\delta=0.032$, $n=1$, corresponding to inclinations of $\sim1.8^\circ$ and eccentricities of $0.032$.  The typical mutual inclination lies firmly in the range $1.0^\circ - 2.2^\circ$, showing that planetary systems tend to be quite flat.

\epsscale{1.0}
\epsscale{1.0}

In contrast to this narrow range of mutual inclination, the $\xi$ distribution can be acceptably matched ($p$-value $>0.01$) over a wide range of eccentricities.  The geometrical reason for the different dependence on inclination and eccentricity is that if inclinations change by 1\%, the duration may change by order unity, but if eccentricities change by 1\%, the duration usually changes by $\sim1$\%.  Good fits to the $\xi$ distribution can be obtained both for circular orbits and for eccentricities in energy equipartition with the mutual inclination (i.e., $\sigma_e = 2 \sigma_i$), and indeed up to several times equipartition.  Therefore the $\xi$ distribution is insensitive to the eccentricity distribution, and our conclusion about mutual inclination does not depend on knowing the eccentricity distribution precisely.

Our conclusion that \Keplers planetary systems are flat supports our inference of Paper I, which used the number of planets of each multiplicity to show that the mutual inclination in systems is typically just a few degrees.  However, it was possible that mutual inclinations are larger than $10^\circ$, provided that planetary systems have many more planets ($\gtrsim10$) than expected.  To rule out this latter possibility, the Radial-Velocity (RV) sample was used to place limits on planet multiplicity \citep{2011TremaineDong, 2012Figueira}, breaking the degeneracy and preferring small planetary inclinations of just a few degrees.  This conclusion requires significant overlap between the current RV sample and the \Kepler sample, which has not been independently demonstrated.  Having reached the same conclusion from the \Kepler sample alone, we have increased confidence that most planetary systems within 0.5 AU of their stars are flat.

By simulating all planet configurations and only comparing the doubly-transiting simulated pairs to the data, our determination of $\sigma_i$ is unbiased.  However, a caveat is that the distribution of inclinations may not be well-characterized by a single Rayleigh distribution, and high-inclination components of the actual distribution would contribute less statistical weight because transits of both planets would be seen only rarely.  Thus, as with all applications of parameter-fitting, the limits given on the parameter $\sigma_i$ hold only to the extent that a member of the family of model distributions describes the actual distribution.  Another caveat is that we have used all pairs of planet candidates throughout, such that the $N$-planet systems are represented more, by a total of $N(N-1)/2$ pairs.  Therefore the architectures of larger-$N$ systems carry more statistical weight in this analysis.  

\section{Comparison to the Solar System} \label{sec:sec_sscomp}

We have described the architecture of a set of multiplanet systems whose gross structure is completely alien to our Solar System.  The sample is dominated by planets with radii between $1-4~R_\oplus$ whose orbital periods are of order 10 days; no such planets exist in the Solar System.  A striking feature of the Solar System is its extreme coplanarity.  This property of exoplanet systems has only started being assessed (Paper I; \citealt{2011TremaineDong, 2012Figueira}).  Perhaps no observation is more crucial for theories of the Solar System's formation in a gaseous disk encircling the proto-Sun.  For exoplanetary systems detected by radial velocity, there is typically no information on the inclination of individual planets, and only weak information (from stability, generally) available regarding their inclination with respect to one another.  With \Keplers transit discoveries, we now have a statistical sample to assess the degree of flatness of extrasolar systems.  

To make a quantitative comparison, we computed the Rayleigh distribution of the mutual inclinations for the Solar System planets a Bayesian technique analogous to that used in \S~\ref{sec:fide}.  We used a uniform prior on $\sigma_i$, and since the allowed region is in each case rather narrow, the results are not sensitive to this prior. There are a total of $N (N-1)/2 = 28$ pairs for the $N=8$ planets.  We used the Keplerian elements at J2000 provided by the JPL Solar System Dynamics website to find the set of 28 mutual inclinations\footnote{http://ssd.jpl.nasa.gov}.    The 95\% credible interval was found to be $\sigma_i = 2.5^\circ |^{+0.6^\circ}_{-0.4^\circ}$.  However, secular evolution changes the orbital orientations on $10^5$~yr timescales.  Using MERCURY, we computed the orbits of the 8 planets for 3 Myr starting at the current epoch, determining their 28 mutual inclinations as a function of time.  The best-fit $\sigma_i$ is $3.1^\circ$ on average, and varies in time with a root-mean square (RMS) of $0.4^\circ$.  So the current epoch has a Rayleigh inclination which is 1.5$\sigma$ lower than the long-term average.  The planet Mercury is well-known as an outlier in inclination, and when this exercise is repeated just with the other 7 planets, the result is $\sigma_i = 1.4^\circ |^{+0.4^\circ}_{-0.2^\circ}$ at the current epoch, and a time-averaged value $2.0^\circ$ with an RMS of $0.3^\circ$ on a 3 Myr timescale.  These values are very similar to the values that we have found for the population of multiply transiting exoplanet systems observed by \Kepler ($\sigma_i = 1.0^\circ - 2.2^\circ$).   

From this 3-Myr integration of the Solar System planets, we also investigated eccentricities, analogously to mutual inclinations.  The time-averaged Rayleigh width is $\sigma_e=0.052$ (compare $\sigma_i = 0.054$~radians) for all 8 planets and a time-averaged $\sigma_e=0.033$ (compare $\sigma_i=0.036$~radians) when excluding the planet Mercury.  The fact that eccentricity and inclination scale together in the Solar System may extend to exoplanetary systems like those \Kepler has discovered.  Our mutual inclination results (\S~\ref{sec:durrat}) suggest that their eccentricities may be small ($e \sim 0.03$), although our transit measurements have not yet probed eccentricity this sensitively.  The radial velocity technique has also discovered many systems of $5-30 M_\oplus$ planets \citep{2011Mayor}, but their eccentricities have not yet been measured this precisely either.  This prediction of small eccentricity for small planets is in contrast to the giant exoplanets found to date, but it may continue the trend that lower mass exoplanets have lower eccentricities \citep{2009Wright}.

Finally, we may ask whether the planets of the Solar System show any resonant structure similar to the \Kepler planets.  The only pair close to a first-order mean-motion resonance is Uranus ($4.0 R_{\oplus}$) and Neptune ($3.9 R_{\oplus}$), whose period ratio is $1.96$.   These values lie near the border of the gap in panel c of figure \ref{fig:pratboth}.  As the origin of this gap remains unclear, it is hard to know whether Uranus and Neptune's period ratio has physical significance.

\section{ Conclusion}
\label{sec:conclude}

Using the B13 catalog, which more than doubles the numbers of multiple planet candidate systems compared to the study of Paper I, we have investigated the architecture of planetary systems anew.  We have shown that the candidates avoid close orbital spacings that would destabilize the orbits of real planets.  From this fact we derived a likely fraction of $85-99\%$ of the candidate pairs are really pairs of planets orbiting the same star.  We found that most planetary systems are not resonant, but the distribution of planet period ratios does show interesting clumping just wide of first-order resonances 2:1 and 3:2, and a gap just interior to them.  It is not yet clear how formation or subsequent evolution produces this pattern.  

The flatness of planetary systems, described based on multiplicity statistics by Paper I, was revisited here based on duration ratio statistics.  We affirm and strengthen the result that pairs of planets tend to be well aligned, to within a few degrees.  This new constraint uses the \Kepler data alone and so is a more direct measurement than had been obtained previously.  

\acknowledgements  Funding for this mission is provided by NASA's Science Mission Directorate.  We thank the entire \Kepler team for the many years of work that is making the \Kepler mission so successful.   We thank Emily Fabrycky, Doug Lin, Man-Hoi Lee, Scott Tremaine, and Tsevi Mazeh for helpful conversations and insightful comments. 
D. C. F. acknowledges support for this work was provided by NASA through Hubble Fellowship grant\#HF-51272.01-A awarded by the Space Telescope Science Institute, which is operated by the Association of Universities for Research in Astronomy, Inc., for NASA, under contract NAS 5-26555.
EA was supported by NSF Career grant 0645416.  E.B.F acknowledges support by the National Aeronautics and Space Administration under grant NNX08AR04G issued through the Kepler Participating Scientist Program, and the Center for Exoplanets and Habitable Worlds, which is supported by the Pennsylvania State University, the Eberly College of Science, and the Pennsylvania Space Grant Consortium.  This material is based upon work supported by the National Science Foundation under Grant No. 0707203.  R.C.M. was support by the National Science Foundation Graduate Research Fellowship under Grant No. DGE-0802270. 

\clearpage
\appendix

\section{Regarding the resonance variable \zz}  \label{sec:zz}

In this appendix we discuss the quantity \zz\ in more detail.  The general form of \zz\ is given by:
\begin{equation}
\zeta_{n,j} = (n+1)\left(\frac{j}{\pp - 1} - {\rm Round}\left[\frac{j}{\pp - 1}\right]\right),
\end{equation}
where $\pp$ is the ratio of the periods of the two planets (always greater than unity), $j$ is the resonance order under consideration, and $n$ is the number of resonance orders that are simultaneously being considered.  This last statement means that the real line is partitioned into non-overlapping neighborhoods around MMRs up to order $n$.  The boundaries between resonances are always defined by resonances of the lowest order not considered.  The motivation for defining this quantity was to provide a means of treating all resonances under study equally, even though their neighborhoods differ in size (approaching zero as the index $j \rightarrow \infty$).

For example, in Paper I and in \S~\ref{sec_pratios} of this work, both first and second order resonances are considered ($n = 2$), and the quantities \zone\ and \ztwo\ (here $\zeta_{2,1}$ and $\zeta_{2,2}$) are given by
\begin{equation}
\zeta_{2,1} = 3 \left( \frac{1}{\pp - 1} - {\rm Round}\left[ \frac{1}{\pp - 1 } \right] \right) \label{eq:zeta21}
\end{equation}
and
\begin{equation}
\zeta_{2,2} = 3 \left( \frac{2}{\pp - 1} - {\rm Round}\left[ \frac{2}{\pp - 1 } \right] \right) , \label{eq:zeta22}
\end{equation}
where \zone\ is applied to those planet pairs that fall into the neighborhoods of the first order resonances and \ztwo\ is applied to the pairs in the neighborhoods of the second order resonances.  The boundaries between these resonance neighborhoods are defined by the intermediate third-order resonances, the lowest-order resonances not considered.

\begin{figure}
\includegraphics{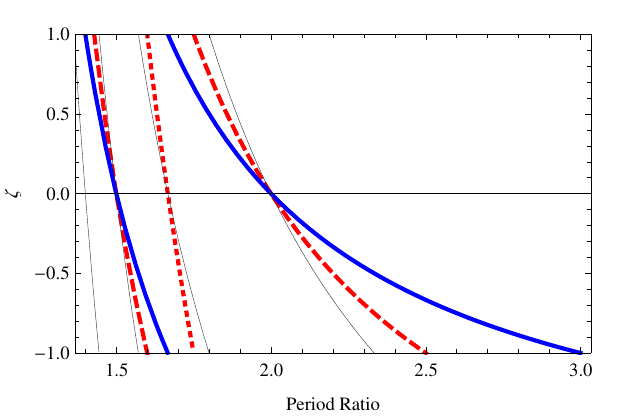}
\caption{The value of \zz\ as a function of the period ratio of two planets.  If only first order resonances are studied, then one uses $\zeta_{1,1}$ (solid, blue) where all period ratios are assigned to a neighborhood of a first-order resonance.  If one simultaneously considers first and second-order resonances, then $\zeta_{2,1}$ (long-dashed, red) and $\zeta_{2,2}$ (short-dashed, red) are used where all period ratios are assigned either to the neighborhood of a first or a second order resonance (these are \zone\ and \ztwo, respectively, of the main text).  Finally, if one wishes to partition the real line into neighborhoods around only second order resonances, then $n=1$ and $j=2$ and the result is $\zeta_{1,2}$, the thin solid curves.\label{zetafig}}
\end{figure}

Suppose, however, that one wanted to assign all period ratios into the neighborhood of a first order resonance only, without considering second order resonances.  Then the proper quantity is $\zeta_{1,1}$, which is contrasted to the $\zeta_{2,1}$ in figure~\ref{zetafig}.  For our sample, choosing such a broad resonance neighborhood includes possible features in the continuum or near the second or higher order resonances and hence dilutes the power of the statistical tests that we employ here.  However, situations may arise where a selection criteria, such as examining only higher-index first order resonances such as 4:3, 5:4, etc., may justify the use of $\zeta_{1,1}$.  Therefore we recommend it for future work with \Keplert, as smaller planets are more likely to be found in such tightly packed configurations, and a longer baseline will have the sensitivity to see them.  To study only second order resonances (including 4:2 and 6:4), one would use the $\zeta_{1,2}$ variable.  Figure \ref{zetafig} contrasts these different choices for mapping period ratios into a space more suitable for studying resonances.

\epsscale{1.1}
\epsscale{1.0}

{\it Facilities:} \facility{Kepler}.

\begin{deluxetable}{cccccccccc}
\tabletypesize{\scriptsize}
\tablecaption{ Characteristics of Planets in Systems with Multiple Transiting Planets \label{tab:multis} }
\tablehead{
\colhead{KOI \#} & 
\colhead{$P$} &
\colhead{$T0$ [BJD]} &
\colhead{$T_{\rm dur}$} &
\colhead{$R_p$} & 
\colhead{S/N} &
\colhead{$M_\star$} &
\colhead{$R_\star$} &
\colhead{$P/P_-$} & 
\colhead{$\Delta_-$} \\
\colhead{ } & 
\colhead{(days)} &
\colhead{$-2454900.0$} &
\colhead{(hr)} &
\colhead{($R_\oplus$)} & 
\colhead{ } &
\colhead{$(M_\odot)$} &
\colhead{$(R_\odot)$} &
\colhead{  } & 
\colhead{  } } 
\startdata
     5.01 &      4.780329 &     65.97325 &    2.0117 &    5.65 &    338.1 &    1.14 &    1.42& & \\
     5.02 &      7.051856 &     66.36690 &    3.6882 &    0.66 &      8.5 &    1.14 &    1.42 &      1.47518 &   8.2\\
    41.02 &      6.887099 &     66.17580 &    4.4764 &    1.23 &     36.8 &    1.10 &    1.23& & \\
    41.01 &     12.815735 &     55.95061 &    6.5383 &    2.08 &     98.5 &    1.10 &    1.23 &      1.86083 &  23.1\\
    41.03 &     35.333143 &     86.98394 &    6.1426 &    1.40 &     23.2 &    1.10 &    1.23 &      2.75701 &  36.0\\
    94.04 &      3.743245 &     64.61390 &    3.6020 &    1.41 &     35.5 &    1.20 &    1.24& & \\
    94.02 &     10.423707 &     71.00718 &    5.2039 &    3.43 &     78.2 &    1.20 &    1.24 &      2.78467 &  28.5\\
    94.01 &     22.343001 &     65.74047 &    6.6986 &    9.25 &    455.3 &    1.20 &    1.24 &      2.14348 &  11.0\\
    94.03 &     54.319931 &     94.23998 &    8.4943 &    5.48 &    206.1 &    1.20 &    1.24 &      2.43118 &  12.0\\
    70.02 &      3.696122 &     67.50026 &    2.4981 &    1.92 &    134.5 &    0.91 &    0.94& & \\
    70.04 &      6.098495 &     68.93378 &    2.7502 &    0.91 &     23.4 &    0.91 &    0.94 &      1.64997 &  19.3\\
    70.01 &     10.854091 &     71.60748 &    3.8004 &    3.09 &    260.2 &    0.91 &    0.94 &      1.77980 &  16.6\\
    70.05 &     19.577893 &     68.20094 &    3.6029 &    1.02 &     18.4 &    0.91 &    0.94 &      1.80373 &  16.8\\
    70.03 &     77.611995 &     97.72630 &    7.2802 &    2.78 &    103.4 &    0.91 &    0.94 &      3.96427 &  39.6\\
   157.06 &     10.304049 &     71.50131 &    4.2052 &    1.89 &     38.6 &    0.98 &    1.06& & \\
   157.01 &     13.024929 &     71.17614 &    4.5733 &    2.92 &     67.6 &    0.98 &    1.06 &      1.26406 &   6.2\\
   157.02 &     22.686708 &     81.45746 &    5.5466 &    3.20 &     73.6 &    0.98 &    1.06 &      1.74179 &  13.4\\
   157.03 &     31.995566 &     87.15961 &    4.3114 &    4.37 &     87.5 &    0.98 &    1.06 &      1.41032 &   7.1\\
   157.04 &     46.687124 &    158.03345 &    6.4943 &    2.60 &     40.5 &    0.98 &    1.06 &      1.45917 &   8.2\\
   157.05 &    118.363821 &    220.31606 &    9.5398 &    3.43 &     54.0 &    0.98 &    1.06 &      2.53526 &  22.1
\enddata
\tablecomments{ Only a portion of this table is shown (one system for each multiplicity), as guidance of its form and content; the entire table is available online. Within each system, the planets are ordered by increasing period.  The $P/P_-$ and $\Delta_-$ columns refer to the spacing between this planet and the next closest planet.  The decimal part of KOI numbers (``.01'', ``.02'', etc.) refers to the order of discovery. }
\end{deluxetable}

\bibliography{msarch} \bibliographystyle{apj}

\end{document}